\documentclass{iaus}
\usepackage{graphicx}

\title[The ARTIST project] 
{Adaptable Radiative Transfer Innovations \\ for Submillimeter Telescopes (ARTIST)}

\author[Marco Padovani \& Jes K. J{\o}rgensen]   
{Marco Padovani$^1$
 \and Jes K. J{\o}rgensen$^2$\\
 on behalf of the ARTIST team:\\[\affilskip]
 Frank Bertoldi$^{3}$,
 Christian Brinch$^{4}$,
 Pau Frau$^{1}$,\\
 Josep Miquel Girart$^{1}$,
 Michiel Hogerheijde$^{4}$, 
 Attila Juhasz$^{4}$,\\
 Rolf Kuiper$^{3}$,
 Reinhold Schaaf$^{\,3}$,  
 Wouter H.~T. Vlemmings$^{3}$}
 
\affiliation{$^1$Institut de Ci\`encies de l'Espai (CSIC-IEEC) \\ 
Universitat Aut\`onoma de Barcelona, Spain
\\ email: {\tt [padovani,girart,frau] @ice.cat} \\[\affilskip]
$^2$Centre for Star and Planet Formation \\
University of Copenhagen, Denmark
\\ email: {\tt jes@snm.ku.dk} \\[\affilskip]
$^{3}$Argelander Insitute for Astronomy \\
University of Bonn, Germany
\\ email: {\tt [wouter,bertoldi,rschaaf,kuiper] @astro.uni-bonn.de}
\\[\affilskip]
$^{4}$Leiden Observatory \\
University of Leiden, the Netherlands 
\\ email: {\tt [michiel,brinch,juhasz] @strw.leidenuniv.nl}
}

\pubyear{2010}
\volume{270}  
\pagerange{1--4}
\jname{Computational Star Formation}
\editors{J. Alves, B. Elmegreen, J.~M. Girart \& V. Trimble, eds.}
\begin{document}

\maketitle

\begin{abstract}
  Submillimeter observations are a key for answering many of the big
  questions in modern-day astrophysics, such as how stars and planets
  form, how galaxies evolve, and how material cycles through stars and
  the interstellar medium. With the upcoming large submillimeter
  facilities ALMA and Herschel a new window will open to study these
  questions.  ARTIST is a project funded in context of the European
  ASTRONET program with the aim of developing a next generation model
  suite for comprehensive multi-dimensional radiative transfer
  calculations of the dust and line emission, as well as their
  polarization, to help interpret observations with these
  groundbreaking facilities.  
\end{abstract}

\firstsection 
\section{Introduction}

The Atacama Large Millimeter Array is the largest ground based project
in astronomy. It will be the world's most powerful instrument for
millimeter and submillimeter astronomy, providing enormous
improvements in sensitivity, resolution and imaging fidelity in these
wavelength bands. The main focus of the early use of ALMA, compared to
present-day facilities, will be its high angular resolution and
sensitivity. For studies of star formation, for example, ALMA will
zoom in to AU scales in circumstellar disks in nearby star forming
regions and thereby address some of the key questions in disk
formation. It will also open up new possibilities in the study of
planet formation, resolving the effects of planets on the disks around
young stars and thus providing direct observational constraints on
planet formation models. At the same time, the Herschel Space
Observatory is providing high spatial and spectral resolution
observations at wavelengths unobservable from the ground, in
particular, of lines of H$_{2}$O and high excitation transitions of
molecules tracing the dense warm gas, e.g., in the innermost regions
of protostars or in shocks.

The simultaneous observation of dust and of a multitude of spectral
lines opens new horizons for the physical and chemical analysis of the
objects, e.g., to determine the excitation conditions (density,
temperature, radiation), the chemical abundances and chemical network,
and through polarization measurements, the magnetic field. It will be
necessary to take all these observational constraints into account for
a realistic quantitative description. ALMA will offer a new chance to
study magnetic fields: due to receiver constraints, full polarization
calibration and imaging will be the norm rather than the exception,
making it essential that a self-consistent polarization modeling tool
is available to the ALMA users.

With the novelty of these observations it will be critical to have an
efficient, flexible and state-of-the-art modeling package that can
provide a direct link between the theoretical predictions and the
quantitative constraints from the submillimeter observations. The new
observational opportunities require a new generation of modeling tools
that can model the full multi-dimensional structure of, e.g., a
low-mass protostar, including its envelope, disk, outflow and magnetic
field, and their time evolution. Current tools are inadequate for the
modeling of such complex structures because of their speed and
inaccessibility, while tools to model polarization are completely
lacking. Both ALMA and Herschel will provide us with large samples of
sources, observed homogeneously as part of large key and legacy
programs. This makes it prudent to have easily accessible and
efficient tools, which with high convergence speed incorporate all
observational constraints, for large source samples, and in a
systematic fashion.  It is the aim of this program to provide such an
innovative suite of model tools and test it with existing
submillimeter data and with new data from ALMA and Herschel.


\section{Objectives}

The goal of this project is to deliver a next generation radiative transfer modeling 
package that provides a self-consistent model for the emission of a multi-dimensional 
source observed at submillimeter wavelengths. For this project we are specifically 
motivated, without any loss of more general applicability, by low-mass star formation. 
Our modeling package shall be able to provide a self-consistent modeling tool for the 
line and continuum as well as polarization emission from, e.g., a young stellar object, 
incorporating an infalling large-scale envelope, rotationally decoupled protoplanetary 
disk, outflow cavity, and magnetic field - with no restrictions of the intrinsic 
geometry. With such an innovative model tool it will be possible to provide quantitative 
constraints on the relation between the large-scale angular momentum in the core and the 
disk evolution, on the direct impact of the outflows and their launching close to the 
disk surface, and on the importance of the magnetic field. Two important issues need to 
be addressed in this particular example:
$(i)$ young stellar objects are characterized by structure on a wide range of spatial 
scales 
and with complex geometries, but 
current radiative transfer tools are locked to fixed linear 
or logarithmic grids and can therefore model multi-dimensional source structures only 
with great computational time expense;
$(ii)$ while most current observations indicate that magnetic fields play an important role in 
various stages of the star formation process, their relation to the physical source 
structure is yet poorly constrained. ALMA observations will study the magnetic fields 
through resolved line and continuum polarization observations that can not be properly 
analyzed with current models.
The modeling tool will also be useful for tackling scientific questions relating to ALMA 
observations of, e.g., evolved stars, planetary nebulae or extragalactic starbursts. 
Providing radiative transfer tools that take complex source structures into account will 
also be of great help to interpret Herschel observations, e.g., to understand the origin 
of H$_{2}$O and high excitation CO lines in the interface between protostellar cores and 
outflows or jets. There is a high degree of coupling between different modeling aspects: 
e.g., to understand the polarization of a given molecule's emission it is necessary to 
understand the physical conditions and chemical network that leads to the molecule's 
formation and excitation. For complex source structures it is necessary to develop an 
approach to time-dependent multi-dimensional modeling, e.g., through libraries of 
theoretical model prescriptions that can readily be incorporated and expanded for 
comparison to the data. 

\section{Tools}
\begin{figure}\centering
\resizebox{0.83\hsize}{!}{\includegraphics[angle=270]{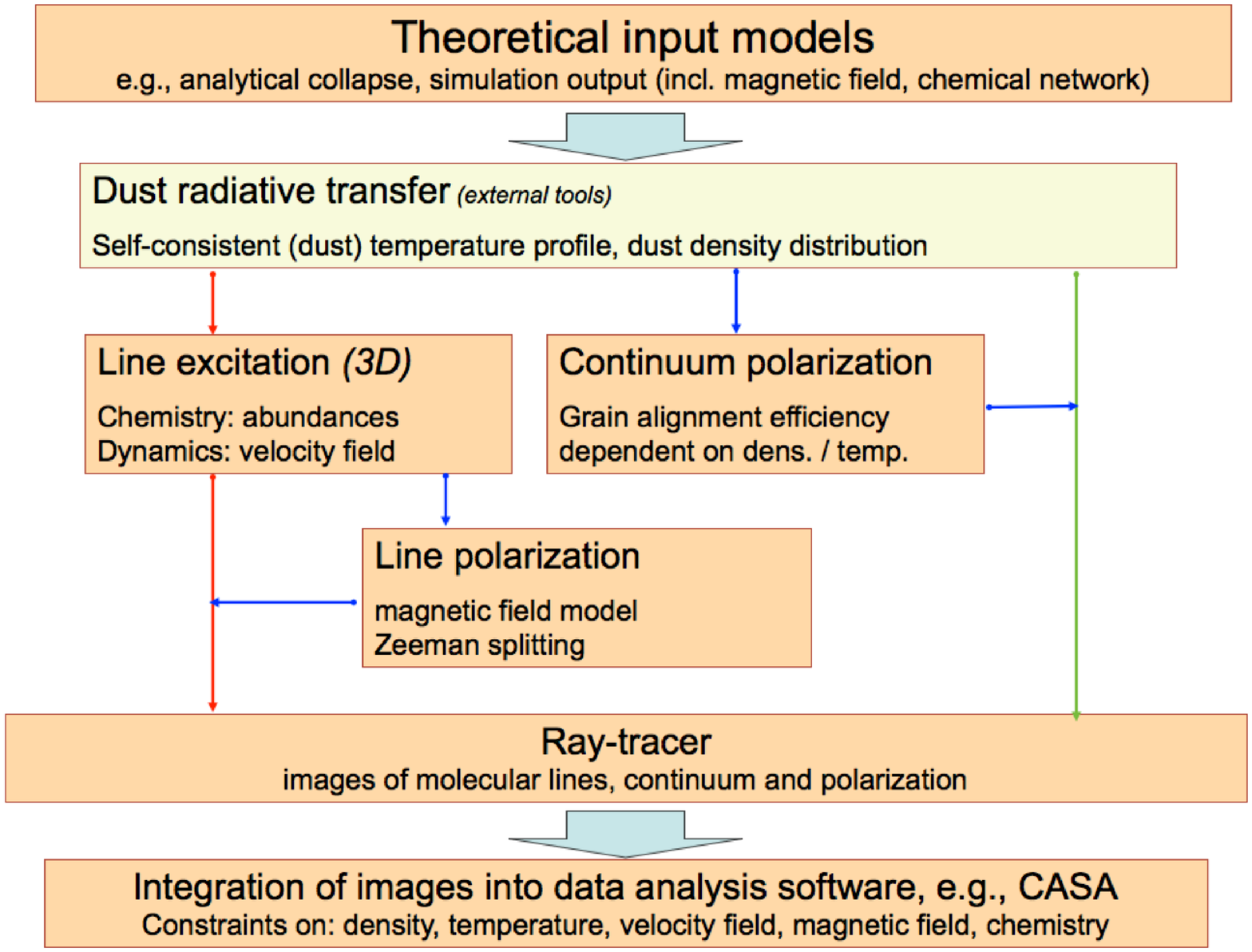}}
\caption{Schematic overview of the components in the ARTIST program.}
\label{schematic}
\end{figure}

The planned model suite will have the following three 
components: $(a)$ an innovative radiative transfer 
code using adaptive gridding that allows simulations of sources with arbitrary 
multi-dimensional and time-dependent structures ensuring a rapid convergence and thus 
allowing an exploration of parameters; 
$(b)$ unique tools for modeling the polarization 
of 
the line and dust emission, information that will come with standard ALMA observations;
$(c)$ a comprehensive Python-based interface connecting these packages, thus with 
direct link to, e.g., ALMA data reduction software (CASA). A schematic
overview of the program is shown in Fig.~\ref{schematic}. \\[-2.0ex]

{\underline{\it (a) $-$ 3D Line Radiative Transfer: LIME.}}
ALMA's high resolution data will produce the need to model phenomena with non-symmetric 
structures, such as spiral-waves, proto-planet resonances in evolving circumstellar 
disks, close protostellar binaries etc.
Conventional radiative transfer tools use simple linear or logarithmic spatial grids.
To  model complex source structures in higher dimensions thus requires increasingly 
finer 
grid scaling, which becomes very difficult to handle computationally.
As an alternative, the SimpleX algorithm (\cite[Ritzerveld \& Icke 2006]{RitzerveldIcke06}) uses a Poisson method 
to define a grid based on the density distribution.
The cornerstone of ARTIST is the 3D line radiative transfer code, LIME 
(Brinch \& Hogerheijde 2010), which utilizes the SimpleX 
gridding algorithm in a 3D 
extension to the RATRAN radiative transfer code 
(\cite[Hogerheijde \& van der Tak 2000]{HogerheijdevanderTak00}). LIME 
is currently being applied to for example modeling new H$_{2}$O observations of 
protostars 
and disks from the Herschel Space Observatory. For these models the adaptive gridding 
method ensures rapid convergence, for lines from molecules
such as H$_{2}$O that are far from 
LTE excitation.\\[-2.0ex]

{\underline{\it (b) $-$ Dust and Line Polarization.}}
Various theoretical studies predict that magnetic fields, turbulence or/and 
magneto-hydrodynamic waves may be the main agents controlling both the evolution of 
molecular 
clouds and the star-formation process (e.g., 
\cite[Bertoldi \& McKee 1992]{BertoldiMcKee92}; 
\cite[Mac Low \& Klessen 2004]{MacLowKlessen04};
\cite[Mouschovias et al. 2006]{Mouschovias06}; 
\cite[van Loo et al. 2007]{vanLoo07}). Unfortunately, the magnetic 
field is the least-known observable in star formation, due to the inherent difficulty to 
measure it with present telescopes (mainly through polarimetric observations of dust and 
molecular emission; e.g., 
\cite[Girart, Rao \& Marrone 2006]{Girartetal06}).
This situation will change dramatically with ALMA, which will provide such an 
improvement in sensitivity that polarization observations of dust and molecular lines	
can be done in many sources at a very good angular resolution. However, the 
interpretation of the polarized data is difficult and appropriate tools are needed to 
scientifically harvest the wealth of polarization data expected from ALMA. Note that the 
ALMA band-7 (275 - 373 GHz) receivers will require full polarization calibration and 
imaging even for projects with no primary interest in polarization.
The ARTIST package consists of a set of modeling tools for polarization and magnetic 
fields in different molecular and dust environments (e.g., low- and high-mass 
protostars, envelopes around evolved stars). \\[-2.0ex]

{\underline{\it (c) $-$ Model interface and library.}}
An important component of ARTIST is a common interface for the codes used for radiative 
transfer modeling of typical data produced by submillimeter telescopes such as ALMA and 
Herschel. The model interface will include: $(i)$ a library of standard input models, 
e.g., for collapsing protostars, circumstellar disks and evolved stars; 
$(ii)$ a wrapper package that links the existing dust and line radiative transfer codes;
$(iii)$ tools to analyze model output, e.g., to extract information about molecular 
excitation and its deviation from LTE, or optical depth surfaces, and to import this in 
existing visualization packages;
$(iv)$ a ray tracing backend that readily provides data cubes that can be used in data 
reduction packages.

\section{Current Status}
The aim of the ARTIST project is to supply the community with the described tools in one 
coherent modeling package. The tools will be made publicly available as they are 
finished: we expect LIME to be released medio-2010 with the remaining components of the 
package to follow. The tools will be distributed and supported through the ALMA 
regional center nodes in Bonn and Leiden as well as the Danish initiative for 
Far-infrared and Submillimeter Astronomy (DFSA; Copenhagen, Denmark). For
more information see {\tt \verb+http://www.astro.uni-bonn.de/ARC/artist+}.


\begin{thebibliography}{}

\bibitem[Bertoldi \& McKee 1992]{BertoldiMcKee92}
{Bertoldi, F. \& McKee, C.~F.} 1992,
\textit{ApJ}, 395, 140

\bibitem[Brinch \& Hogerhejiijde 2010]{Brinch10}
{Brinch, C. \& Hogerheijde, M.} 2010,
\textit{A\&A}, 523, 25

\bibitem[Girart, Rao \& Marrone 2006]{Girartetal06}
{Girart, J.~M., Rao, R. \& Marrone, D.~P.} 2006
\textit{Science}, 313, 812

\bibitem[Hogerheijde \& van der Tak 2000]{HogerheijdevanderTak00}
{Hogerheijde, M.~M. \& van der Tak, F.~F.~S.} 2000,
\textit{A\&A}, 362, 697

\bibitem[Mac Low \& Klessen 2004]{MacLowKlessen04}
{Mac Low, M.-M. \& Klessen, R.~S.} 2004,
\textit{RvMP}, 76, 125

\bibitem[Mouschovias et al. 2006]{Mouschovias06}
{Mouschovias, T.~Ch., Tassis, K. \& Kunz, M.~W.} 2006,
\textit{ApJ}, 646, 1043

\bibitem[Ritzerveld \& Icke (2006)]{RitzerveldIcke06}
{Ritzerveld, J. \& Icke, V.} 2006,
\textit{PhRvE}, 74, 26704

\bibitem[van Loo et al. 2007]{vanLoo07}
{van Loo, S., Falle, S.~A.~E.~G., Hartquist, T.~W. \& Moore, T.~J.~T.} 2007
\textit{A\&A}, 471, 213

\end{thebibliography}
\end{document}